\documentclass[submission,creativecommons,copyright,sharealike,noncommercial]{eptcs}
\providecommand{\event}{TARK 2023} 

\usepackage{iftex}

\ifpdf
  \usepackage{underscore}         
  \usepackage[T1]{fontenc}        
\else
  \usepackage{breakurl}           
\fi

\title{An Abstract Look at\\Awareness Models and Their Dynamics}
\author{Carlo Proietti
\institute{ILC, CNR\\ Genova, Italy}
\email{carlo.proietti@ilc.cnr.it}
\and
Fernando R. Vel\'azquez-Quesada
\institute{University of Bergen\\
Norway}
\email{Fernando.VelaquezQuesada@uib.no}
\and
Antonio Yuste-Ginel \qquad\qquad
\institute{Universidad Complutense de Madrid,\\
Spain \thanks{The research activity of Antonio Yuste-Ginel was partly funded by the project PID2020-117871GB-I00 whose principal investigator is Francisco Salguero Samillar. Part of the research activity that crystallised in this paper was carried out when he was employed at the IRIT (Toulouse) by a postdoc contract funded by EUICT-482020 project TAILOR GA952215.}}
\email{antoyust@ucm.es}
}
\def\titlerunning{An Abstract Look at Awareness Models and Their Dynamics}
\def\authorrunning{C. Proietti, F.R. Vel\'azquez-Quesada \& A. Yuste-Ginel}

\usepackage[cmex10]{amsmath}
\usepackage{scalerel}
\usepackage{amssymb}
\usepackage{latexsym}
\usepackage{mathtools}
\usepackage{color}
\usepackage{hyperref}
\usepackage{underscore}
\usepackage{booktabs}
\usepackage{comment}
\usepackage{xspace}
\usepackage{lineno}
\usepackage{multicol}
\usepackage[mathcal]{euscript}

\usepackage[UKenglish]{babel}

\usepackage{tikz}
\usetikzlibrary{positioning,arrows,calc,shapes,babel,fit}

\tikzset{
modal/.style={>=stealth',shorten >=1pt,shorten <=1pt,auto,node distance=1.5cm,
semithick},
event/.style={draw,minimum size=0.5cm,fill=gray!15},
world/.style={draw,minimum size=0.5cm,fill=gray!15},
point/.style={circle,draw,inner sep=0.5mm,fill=black},
reflexive above/.style={->,loop,looseness=4,in=120,out=60},
reflexive below/.style={->,loop,looseness=4,in=240,out=300},
reflexive left/.style={->,loop,looseness=4,in=150,out=210},
reflexive right/.style={->,loop,looseness=4,in=30,out=330}
}

\usepackage[textsize=scriptsize]{todonotes}

\definecolor{colorfer}{RGB}{255,96,0}
\definecolor{colorcarlo}{RGB}{142, 209, 144}
\definecolor{coloranto}{RGB}{214, 108, 143}
\newcommand{\notaanto}[1]{\todo[color=coloranto!15]{\textbf{Anto}: #1}}
\newcommand{\notafer}[1]{\todo[color=colorfer!15]{\textbf{Fer}: #1}}

\usepackage[hyperref,thmmarks]{ntheorem}

\theoremstyle{plain}
\theorembodyfont{\itshape}
\theoremsymbol{\ensuremath{\blacksquare}}

\newtheorem{theorem}{Theorem}
\newtheorem{proposition}{Proposition}

\newtheorem{rem}{Remark}

\theoremheaderfont{\normalfont\bfseries}
\theorembodyfont{\itshape}
\theoremsymbol{\ensuremath{\blacktriangleleft}}

\newtheorem{definition}{Definition}
\newtheorem{example}{Example}

\theoremheaderfont{\itshape}
\theorembodyfont{\normalfont}
\theoremstyle{nonumberplain}
\theoremsymbol{\ensuremath{\blacksquare}}

\newtheorem{proof}{Proof.}

\usepackage{enumitem}

\newcommand{\itmformatstyle}{\bfseries \itshape}

\setlist[enumerate,1]{label={\itmformatstyle (\roman*)}}
\setlist[enumerate,2]{label*={\itmformatstyle (\alph*)}}

\newlist{inlineenum}{enumerate*}{1}
\setlist[inlineenum,1]{label={\itmformatstyle (\roman*)}}

\newlist{compactenumerate}{enumerate}{1}
\setlist[compactenumerate,1]{label={\itmformatstyle (\roman*)}, leftmargin=1.5em, rightmargin=0em, topsep=0.25em, itemsep=0.25em}

\setlist[itemize,1]{label=\textbullet}
\setlist[itemize,2]{label=--}

\newlist{compactitemize}{itemize}{3}
\setlist[compactitemize,1]{label=\textbullet, leftmargin=1.5em, rightmargin=0em, topsep=0.25em, itemsep=0.25em}
\setlist[compactitemize,2]{label=--, leftmargin=1.25em, rightmargin=0em, topsep=0.25em, itemsep=0.25em}

\addto\extrasUKenglish{%

}

\newcommand{\autorefp}[1]{(\autoref{#1})}

\newenvironment{ctabular}[1]
{\begin{center}\begin{tabular}{#1}}
{\end{tabular} \end{center}}

\newenvironment{smallctabular}[1]
{\begin{center}\begin{small}\begin{tabular}{#1}}
{\end{tabular}\end{small}\end{center}}

\newcommand{\nsparagraph}[1]{\noindent\textbf{#1}\,}
\newcommand{\msparagraph}[1]{\medskip\noindent\textbf{#1}\,}

\newcommand{\imply}{\,\Rightarrow\,}
\newcommand{\atm}{\operatorname{atm}}
\newcommand{\set}[1]{\left\{ #1 \right\}}

\newcommand{\classobmod}{\mathfrak{M}^{\obj}}

\newcommand{\nmod}{\mathcal{M}}
\newcommand{\wset}{\mathcal{W}}
\newcommand{\acc}{\mathcal{R}}
\newcommand{\valfun}{\mathcal{V}}

\newcommand{\lanset}{\mathcal{L}}
\newcommand{\lansetjust}{\mathcal{L}_J}

\newcommand{\obj}{\mathbf{O}}
\newcommand{\at}{\mathsf{At}}
\newcommand{\ag}{\mathsf{Ag}}

\newcommand{\obfun}{\mathcal{O}}
\newcommand{\awfun}{\mathcal{A}}
\newcommand{\jusfun}{\mathcal{E}}

\newcommand{\objmod}{(\wset, \acc, \obfun, \valfun)}

\newcommand{\modawa}{\operatorname{A}}
\newcommand{\modawag}{\widetilde{\modawa}}
\newcommand{\modobj}{\operatorname{O}}

\newcommand{\emp}{\mathsf{EMP}}
\newcommand{\nprop}{\mathbb{P}}

\newcommand{\nemod}{\mathcal{E}}
\newcommand{\emod}{(\eveset, \T, \pre, \eff)}
\newcommand{\eveset}{\mathcal{S}}
\newcommand{\T}{\mathcal{T}}
\newcommand{\pre}{\mathsf{pre}}
\newcommand{\eff}{\mathsf{eff}}

\makeindex

\newcommand{\kripke}{\mathsf{K}}
\newcommand{\nlog}{\mathsf{L}}
\begin{document}
\maketitle
%
%
\begin{abstract}
This work builds upon a well-established research tradition on modal logics of awareness. One of its aims is to export tools and techniques to other areas within modal logic. To this end, we illustrate a number of significant bridges with abstract argumentation, justification logics, the epistemic logic of knowing-what and deontic logic, where basic notions and definitional concepts can be expressed in terms of the awareness operator combined with the $\Box$ modality. Furthermore, these conceptual links point to interesting properties of awareness sets beyond those standardly assumed in awareness logics – i.e. positive and negative introspection. We show that the properties we list are characterised by corresponding canonical formulas, so {as} to obtain a series of off-the-shelf axiomatisations for them. As a second focus, we investigate the general dynamics of this framework by means of event models. Of specific interest in this context is to know under which conditions, given a model that satisfies some property, the update with an event model keeps it within the intended class. This is known as the closure problem in general dynamic epistemic logics. As a main contribution, we prove a number of closure theorems providing sufficient conditions for the preservation of our properties. Again, these results enable us to axiomatize our dynamic logics by means of reduction axioms.
\end{abstract}

\sloppy
\section{Introduction}\label{sec:intro}

Epistemic logics of awareness \cite{FH1987,Schipper2015} are extensions of propositional epistemic logic (EL; \cite{hintikka1962knowledge}) introduced for modelling a form of (\textit{explicit}) knowledge that lacks closure under logical consequence (therefore avoiding the \textit{logical omniscience} problem). The idea is that knowledge requires both lack of uncertainty (the standard $\Box$ modality) \emph{and} awareness, with the latter a unary modality that, semantically, verifies whether the given formula belongs to a specified world-dependant \emph{awareness} set. One can deal with specific awareness properties (e.g., awareness introspection) by specifying not only the properties of the awareness sets but also their interaction with the accessibility relations. One can also look at dynamics of awareness in the dynamic epistemic logic style (DEL; \cite{baltag1998logic,hans2007,van2011book,sep-dynamic-epistemic}), defining model-changing actions for representing acts of awareness \textit{elicitation} or \textit{forgetting} \cite{van2010dynamics,velazquez2010small,van2012action,GrossiVelazquez2015}. \par
The epistemic awareness setting can also be interpreted more generally by abstracting away from this specific reading (see \autoref{sec:basic}).
At a general level, one can read the awareness entities as a set $\obj$ of generic objects, and the corresponding awareness modality as capturing the notion of ``owning some abstract object $o \in \obj$''. By doing so, one can find connections with other modal logics where abstract objects are used as additional or definitional concepts. For example, other approaches in epistemic and deontic logic \cite{plaza1989,kanger1970new,hansson1969analysis,van2014priority} can be seen as instances of a more general awareness-like framework. From this perspective, model properties connecting the ``owning-the-object'' operator ``$\modobj$'' with $\Box$ constitute interesting desiderata. This paper defines a number of such properties and characterises them with formulas of the $\obj$-language. \par 
A second aim of this work is to investigate the dynamics of general $\obj$-models. We use \textit{event models} as in \cite{baltag2004logics} for their power to encode epistemic and factual changes at an extreme level of granularity \cite{van2008semantic}. Yet, a drawback of it is the often non-trivial \textit{closure problem}: guaranteeing that, for a given class $\mathfrak{M}$ of models, the product update of an $\mathfrak{M}$-model with an event model remains in $\mathfrak{M}$. Closure results clarify the general constraints for the executability of actions, and therefore provide safe guidance for modelling them. Some closure theorems are available for DEL, establishing sufficient conditions for the preservation of accessibility relations \cite{aucher2008consistency,sep-dynamic-epistemic}. However, this issue is relatively underexplored for properties relating accessibility relations and awareness sets, as the ones mentioned above (with the exception of \cite{synthese}). As a central contribution of our work, we prove closure theorems for these properties. As an important byproduct, this serves to find direct roads to axiomatisation via reduction axioms. \par 

The paper proceeds as follows. \autoref{sec:basic} introduces the general $\obj$-framework, illustrating some of its applications. Crucially, it also lists meaningful model properties (at both the individual and multi-agent level; \autoref{sbs:properties}), providing their syntactic characterizations as well as their complete axiomatisations as a main result. \autoref{sec:dynamics} is about the dynamics of $\obj$-models, semantically: we introduce event models and the closure problem, identifying sufficient conditions for the {preservation} of the discussed model properties. \autoref{sec:axioms} looks at dynamics from the syntactic side, providing sound and complete axiomatisations for dynamic $\obj$-logics. We end with a discussion of our results in \autoref{sec:appendix}. Sketches of proofs are to be found in the \nameref{sec:appendix}.


\section{Basic framework}\label{sec:basic}

Through this document, let $\ag$ be a finite non-empty set of agents, $\at$ be a countable set of propositional variables, and $\obj$ be a countable non-empty set of abstract objects. An $\obj$-model is just a multi-relational model together with a function that assigns, to each agent, a set of objects from $\obj$ at each possible world. 
\begin{definition}[$\obj$-Model]
  An \emph{$\obj$-model} is a tuple $\nmod = \objmod$ where $\wset \neq \emptyset$ is a set whose elements are called possible worlds,
  $\acc: \ag \to \wp(\wset \times \wset)$ assigns a binary relation on $\wset$ to each agent $i \in \ag$, $\obfun: (\ag \times \wset) \to \wp(\obj)$ assigns a set of objects to each agent $i \in \ag$ at each world $w \in \wset$, and $\valfun:\at \to \wp(\wset)$ is an atomic valuation function. Note: $\acc_i$ abbreviates $\acc(i)$, and $\obfun_i(w)$ stands for $\obfun(i,w)$. The set of worlds of a given $\nmod$ is referred to as $\wset[\nmod]$; the same convention applies $\acc$, $\obfun$ and $\valfun$. We use infix notation for each $\acc_i$. A \emph{pointed $\obj$-model} is a tuple $(\nmod, w)$ with $\nmod$ an $\obj$-model and $w \in \wset[\nmod]$. Finally, $\classobmod$ denotes the class of all $\obj$-models.
\end{definition}

The language for describing $\obj$-models is the following.

\begin{definition}[Language $\lanset$]
  Given $\ag$, $\at$, and $\obj$ as above, formulas $\varphi$ of the language $\lanset$ are given by
  \begin{ctabular}{c}
    $\varphi ::= \top \mid p \mid \modobj_i o \mid \lnot \varphi \mid \varphi \land \varphi \mid \square_i \varphi$
  \end{ctabular}
  with $p \in \at$, $i \in \ag$ and $o \in \obj$. Other Boolean constants/operators are defined as usual; likewise for the modal dual $\Diamond_i \varphi$, defined as $\lnot \square_i \lnot \varphi$. Formulas of $\lanset$ are interpreted at pointed $\obj$-models. The truth-clauses for the multi-modal fragment of $\lanset$ are the standard ones; for the new formulas,
  \begin{ctabular}{r@{\quad\text{iff}\quad}l}
    $\nmod,w \models \modobj_i o$ & $o \in \obfun_i(w)$.
  \end{ctabular}
  Global truth of a formula and a set of formulas in a model is defined as usual \cite{blackburn2002}, and denoted $\nmod\models \varphi$ and $\nmod \models \Phi$, respectively. Likewise for the notion of validity (notation: $\models \varphi$).
\end{definition}

Let us now present some particular interpretations and instantiations of $\obj$-models.

\msparagraph{{Models for general and atomic awareness}} A model for \emph{general} awareness \cite{FH1987} is an $\obj$-model where $\obj$ is the language $\lanset$ itself. In this context, $\obfun_i$ is called the \emph{awareness function} and it is denoted by $\awfun_i$; notationally, 
 the operator $\modobj_i$ is replaced by the awareness operator $\modawa_i$.
A model for {\emph{atomic awareness}} \cite{FH1987,halpern2001alternative} is instead one where the awareness function $\awfun$ returns a set of atoms from $\at$, with agent $i$ aware of $\varphi$ at a world $w$ if and only if the set of atoms in $\varphi$ is a subset of $\awfun_i(w)$. These structures correspond to $\obj$-models where $\obj$ is a set of atoms $\at$. Syntactically, $\nmod,w \models \modawa_i p$ iff $p \in \awfun_i(w)$, and then one can define inductively an additional modality $\modawag_i$ that works over arbitrary formulas:
\begin{ctabular}{ccc}
  \begin{tabular}{r@{\;:=\;}l}
    $\modawag_i \top$        & $\top$, \\
    $\modawag_i p$           & $\modawa_i p$, \\
    $\modawag_i \modawa_j p$ & $\modawag_i p$.
  \end{tabular}
  &
  \begin{tabular}{r@{\;:=\;}l}
    $\modawag_i \lnot \varphi$        & $\modawag_i \varphi$, \\
    $\modawag_i (\varphi \land \psi)$ & $\modawag_i \varphi \land \modawag_i \varphi$, \\
    \multicolumn{2}{c}{}
  \end{tabular}
  &
  \begin{tabular}{r@{\;:=\;}l}
    $\modawag_i \square_j \varphi$  & $\modawag_i \varphi$, \\
    $\modawag_i \modawag_j \varphi$ & $\modawag_i \varphi$. \\
    \multicolumn{2}{c}{}
  \end{tabular}
\end{ctabular}
In this way, $\nmod,w \models \modawag_i \varphi$ if and only if $\atm(\varphi) \subseteq \awfun_i(w)$, with $\atm(\varphi)$ the set of atoms occurring in $\varphi$.

\msparagraph{Models for awareness of arguments} One can also conceive $\obj$ as a set of \emph{abstract arguments} and $\obfun$ as a function indicating the set of arguments that each agent is aware of at each world \cite{schwarzentruber2012building}, {so that $\obj_i a$ means ``agent $i$ is aware of argument $a$'' or ``agent $i$ is able to use argument $a$''}. The resulting models constitute `epistemic' versions of the abstract models of argumentation introduced in \cite{dung1995acceptability}. The main idea behind \textit{abstract argumentation} is to represent arguments as nodes of a graph, and attacks among them as arrows of the graph. There, argumentative notions such as argument acceptability are reduced to graph-theoretical notions, such as stability of a set within a graph. In the modalized (multi-agent) versions, a possible world is constituted by one such graph plus the specification of which arguments and attacks each agent is aware of. This enables us to express higher-order uncertainty about awareness of arguments \cite{schwarzentruber2012building}, which is in turn crucial for modelling strategic reasoning in an argumentative environment \cite{thimm2014strategic} and its dynamics \cite{dali,synthese}.
In a similar vein, $\obj$-models have been applied to more structured frameworks for argumentation \cite{comma,tark}, with $\obj$ understood as a set of ASPIC$^{+}$ arguments \cite{modgil2013}.

\msparagraph{Justification logics} In the justification logics of \cite{artemov-fitting-sep}, justifications are abstract objects which have structure and operations on them. Formally, the set of \emph{justification terms} $J$ is built from sets of justification constants and justification variables by means of the operations of application (`$\cdot$') and sum (`$+$'). Thanks to them, one can define the language $\lansetjust$ as the basic (multi)modal language plus expressions of the form $t{:}\varphi$ (with $t$ a term and $\varphi$ a formula), read as \emph{``$t$ is a justification for $\varphi$''}. Formulas of this extended language are interpreted over \emph{justification models}, tuples $M = (\wset, \acc, \jusfun, \valfun)$ where $\wset$, $\acc$ and $\valfun$ are as in a $\obj$-models. The new component, the evidence function $\jusfun:(J \times \lansetjust) \to \wp(\wset)$, provides the set of worlds $\jusfun(t,\varphi)$ in which the term $t$ is relevant/admissible evidence for the formula $\varphi$. For this to work properly, $\jusfun$ should satisfy both
\[
  \jusfun(s,\varphi \to \psi) \cap \jusfun(t,\varphi) \subseteq \jusfun(s{\cdot}t,\psi)
  \quad\text{and}\quad
  \jusfun(s,\varphi) \cup \jusfun(t,\varphi) \subseteq \jusfun(s+t,\varphi).
\]
Then, $(M, w) \models t{:}\varphi$ if and only if both $w \in \jusfun(t,\varphi)$ and $\varphi$ holds in all worlds $\acc$-reachable from $w$.

\smallskip

A justification model can be seen as an $\obj$-model in which the codomain of $\obfun$ is a set of pairs of the form (justification, formula). Indeed, the evidence function can be equivalently defined as $\jusfun':\wset \to \wp(J \times \lansetjust)$, with $(t,\varphi) \in \jusfun(w)$ indicating that $t$ is relevant/admissible for $\varphi$ at $w$. Its constraints become
\[
  \set{(s,\varphi \to \psi), (t,\varphi)} \subseteq \jusfun(w) \imply (s{\cdot}t,\psi) \in \jusfun(w)
  \quad\text{and}\quad
  (s,\varphi) \in \jusfun(w) \imply \set{(s+t,\varphi), (t+s,\varphi)} \subseteq \jusfun(w)
\]
and thus a justification model $M=(\wset, \acc, \jusfun, \valfun)$ can be equivalently stated as $M'=(\wset, \acc, \jusfun', \valfun)$. Finally, for the language, one can simply define $t{:}\varphi := \modobj (t,\varphi) \land \Box\varphi$.

\msparagraph{Models for knowing-what} Plaza's analysis of the \emph{knowing-what-the-value-of-a-constant-is} operator (\emph{knowing-what} for short; \cite{plaza1989}) has played a crucial role in the emerging of a \emph{new generation of epistemic logics} \cite{Wang2018} that go beyond standard \emph{knowing-that} modalities. Adding $\mathsf{D}$ as a denumerable set of constants (rigid designators) to the framework, a $\mathsf{D}$-model extends a multi-relational model with a function $\valfun_\mathsf{D}: (\wset \times \mathsf{D})\to S$, assigning a value in $S$ to each object in $\mathsf{D}$ at each world in $\wset$. Syntactically, the language extends the standard modal language with expressions of the form $Kv_i d$ (for $i \in \ag$ and $d \in \mathsf{D}$), intuitively read as ``agent $i$ knows the value of constant $d$''. Semantically, this is the case iff $d$ denotes the same object in all $i$'s epistemically accessible worlds:
\begin{ctabular}{r@{\quad\text{iff}\quad}l}
  $M,w \models_v Kv_{i} d$ & $\forall u,u'\in \wset$, $w\acc_i u$ and $w\acc_i u'$ imply $\valfun_\mathsf{D}(u,d)=\valfun_\mathsf{D}(u',d)$.
\end{ctabular}
A $\mathsf{D}$-model can be seen as an $\obj$-model where $\obj$ is the set of tuples $D \times S$, and with each possible world $w$ having a single set $\obfun(w)$.
\footnote{Alternatively, all $\obfun_i$-sets are the same at each possible world.} Moreover, these sets should contain exactly one pair $(d,s)$ for each $d \in \mathsf{D}$. Finally, using the `owning' operator $\modobj$, the formula $Kv_i d$ is definable as $\mathit{Kv}_{i} d := \Diamond_i \modobj (d,s) \to \square_i \modobj (d,s)$.

\msparagraph{Deontic logic} The \textit{Kanger-Anderson reductionist approach} to deontic logic \cite{kanger1970new, anderson1958reduction} consists in expressing the $OB$ operator `it is obligatory that' by means of the alethic modality $\square$ plus a new propositional constant. In Kanger's terms, {the} propositional constant $d$ has the intuitive meaning `all normative demands are satisfied' (i.e., the situation is `ideal'). The $OB \varphi$ operator is defined as $\Box(d \to \varphi)$, and Kanger’s system of deontic logic is obtained by adding, to the modal logic $K$, the axiom $\Diamond d$, which semantically defines \textit{strong seriality}: for any world $w$ there is a $v$ s.t. $w\acc v$ and $v$ is ideal. From our perspective, it is natural to interpret $\obj$ as representing the set of normative demands. Interestingly, when $\obj$ is finite, it is easy to rewrite $d$ as $\bigwedge_{o \in \obj} o$ and capture its intended meaning. Indeed, the following holds:
\begin{rem}
In the class of $\obj$-models with $\obj$ finite, the formula $\Diamond \bigwedge_{o \in \obj} o$ {characterizes} strong seriality.
\end{rem}
While the original Kanger-Anderson's framework cannot handle contrary-to-duty obligations, further refinements, dating back to \cite{hansson1969analysis}, allow this. The key idea is to use the $\Diamond$ operator to express betterness as a pre-order among worlds, where $\Diamond \varphi$ means that $\varphi$ is the case in some world that is at least as good as the actual. As suggested by \cite{van2014priority}, it is also natural to encode betterness by syntactic means, i.e.\, via an ordering $\prec$ between formulas, where if $\varphi \prec \psi$ then $\psi$ logically implies $\varphi$. Along similar lines, by regarding our objects as normative demands (desirable properties), one can define a betterness ordering as, e.g., $\bigwedge_{o \in S} o \prec \bigwedge_{o \in S'} o $ iff $S \subsetneq S'$, where $S,S' \subseteq \obj$, and therefore $\bigwedge_{o \in \obj} o$ is the maximal element.
\begin{rem}
Under this reading, the formula  $\obj o \to \square \obj o$ characterizes the fact that $\acc$ is a betterness relation: only worlds that are at least as ideal can be seen. Further, $\lnot \obj o \to \lozenge  \obj o$ says that all non-ideal worlds failing some normative demand have access to some world satisfying it. Together with $\obj o \to \square \obj o$ and the axiom $\Diamond \Diamond p \to \Diamond p$ for transitivity, this implies strong seriality.
\end{rem}

\subsection{Some useful/important properties of \texorpdfstring{$\obj$}{O}-models}\label{sbs:properties}

Depending on the particular interpretation, an $\obj$-model may be asked to satisfy requirements
connecting the $\obfun$-sets with the accessibility relations $\acc_i$. This section lists some examples, providing their syntactic characterisations and discussing the settings in which they might be useful/important. \par 
\msparagraph{Individual properties} We start with the simplest properties relating accessibility relations with objects: those whose formulations involve a single agent. These \emph{individual properties} are summarised in \autoref{tbl:properties}, with a model $\nmod$ satisfying an individual property (e.g., preservation of $\obfun$) iff it satisfies it for every agent $i\in \ag$. 
\textit{Preservation} and \textit{anti-preservation} come from awareness logic \cite{FH1987,heifetz2006interactive,Schipper2015}, where they capture the idea of \emph{awareness introspection}. Indeed, if $\acc_i$ preserves (anti-preserves) $\obfun_i$, then agent $i$'s awareness is positively (negatively) introspective: whenever she is (not) aware of something, she knows/believes so. The \textit{invariance} property, the conjunction of preservation and anti-preservation, captures perfect/total awareness introspection. Finally, the \textit{inversion} properties are mathematical variations of the preservation properties: they ask for the accessibility relation to \emph{invert} the `opinion' of a set towards an object. To the best of our knowledge, none of them has been studied, and yet they can be seen as intuitively appealing in some contexts. For instance, $\acc$ inverting $\obfun$ seems appropriate to talk, in the spirit of \cite{anderson1958reduction}, about normative violations in a deontic reading of $\obj$-models: if an agent has a bad habit, then she would prefer not to have it. Analogously, $\acc$ anti-inverting $\obfun$ works well as a formal property for normative demands as those of \cite{kanger1970new}: if the agent lacks it, then she prefers to have it. 
\par \medskip
The following proposition states the definability of the listed individual properties in $\lanset$.

\begin{proposition}\label{pro:individual-formulas}
  Let $\nprop$ be an individual property (left-hand column of \autoref{tbl:properties}); let $\Gamma({\nprop})$ be the set of all instances of the corresponding schema in the right-hand column. For any $\obj$-model $\nmod$, we have that
  \begin{center}
    $\nmod$ satisfies $\nprop$ \quad iff \quad  $\nmod \models \Gamma({\nprop})$.
  \end{center}
\end{proposition}

\begin{table}[t]
  \renewcommand{\arraystretch}{1.2}
  \begin{smallctabular}{lll}
    \toprule
    \textbf{$\objmod$ is s.t.}
    &
    \textbf{iff, for every $w,u \in \wset$,}
    &
    \textbf{Characterising schema}
    \\
    \midrule
    $\acc_i$ preserves $\obfun_i$          & $w \acc_i u \imply \obfun_i(w) \subseteq \obfun_i(u)$     & $\modobj_i o \to \square_i \modobj_i o$ \\
    $\acc_i$ anti-preserves $\obfun_i$     & $w \acc_i u \imply \obfun_i(u) \subseteq \obfun_i(w)$     & $\lnot \modobj_i o \to \square_i \lnot \modobj_i o$ \\
    $\obfun_i$ is invariant under $\acc_i$ & $w \acc_i u \imply \obfun_i(w) = \obfun_i(u)$             & $(\modobj_i o \to \square_i \modobj_i o) \land (\lnot \modobj_i o \to \square_i \lnot \modobj_i o)$ \\
    $\acc_i$ inverts $\obfun_i$            & $w \acc_i u \imply \obfun_i(w)\cap \obfun_i(u)=\emptyset$ &  $\modobj_i o \to \square_i \lnot \modobj_i o$ \\
    $\acc_i$ anti-inverts $\obfun_i$       & $w \acc_i u \imply \obfun_i(w)\cup \obfun_i(u)=\obj$      & $\lnot \modobj_i o \to \square_i \modobj_i o$ \\
    \bottomrule
  \end{smallctabular}
  \caption{Some individual properties.}
  \label{tbl:properties}
\end{table}

\msparagraph{Group properties} These properties express how the set of objects of one agent `affects'/`influences' the set of objects of other agents in the worlds accessible to the first. As it is explained below, the notion of ``a model $\nmod$ satisfying a group property $\nprop$'' should be parametrised to avoid trivialisations (e.g., all agents are aware of everything). The properties are listed in \autoref{tbl:groproperties}, with $f:\ag \to \wp(\ag) \setminus \{\emptyset\}$ a possibly partial function whose domain is non-empty. If $\nprop$ is a group property, we say that \emph{$\nmod$ $f$-satisfies $\nprop$} iff for every $i\in Dom(f)$, $\nmod$ satisfies $\nprop$ for $i$ and $f(i)$. Moreover, we call \emph{universal (resp.\ existential) group properties} those that contain ``for all'' (resp.\ ``for some'') in their formulation. Regarding their use, the property of anti-preservation of $\obj$ for everyone in $f(i)$ was first brought up by \cite{schwarzentruber2012building} in the context of epistemic logics for abstract argumentation: if the agent is not aware of an argument, she thinks no one else is. As suggested by \cite{synthese}, this property makes general sense under a \emph{de re} reading of the epistemic possibility of attributing someone else a given item. The remaining versions of preservation and anti-preservation are natural mathematical variations of the first, and it is not difficult to find intuitive readings for them. 
For instance, in an awareness context, preservation \emph{for all} indicates that each agent $i$ knows/believes that everybody in $f(i)$ is aware of what she is aware of. Analogously, preservation \emph{for some} indicates that each agent $i$ knows/believes that at least someone in $f(i)$ is aware of what she is aware of.

\begin{table}[t]
  \renewcommand{\arraystretch}{1.2}
  \begin{smallctabular}{@{}l@{\;\;}l@{\quad}l@{}}
    \toprule
    \textbf{$\objmod$ is s.t.}
    &
    \textbf{iff, for every $w,u \in \wset$,}
    &
    \textbf{Characterising schema}
    \\
    \midrule
    $\acc_i$ preserves $\obfun_j$ for all $j \in f(i) \subseteq \ag$       & $w \acc_i u \imply \obfun_i(w) \subseteq \bigcap_{j \in f(i)} \obfun_j(u)$ & $\modobj_i o \to \square_i \bigwedge_{j \in f(i)} \modobj_j o$ \\
    $\acc_i$ preserves $\obfun_j$ for some $j \in f(i) \subseteq \ag$      & $w \acc_i u \imply \obfun_i(w) \subseteq \bigcup_{j \in f(i)} \obfun_j(u)$ & $\modobj_i o \to \square_i \bigvee_{j \in f(i)} \modobj_j o$ \\    
    $\acc_i$ anti-preserves $\obfun_j$ for all $j \in f(i) \subseteq \ag$  & $w \acc_i u \imply \bigcup_{j \in f(i)} \obfun_j(u) \subseteq \obfun_i(w)$ & $\lnot \modobj_i o \to \square_i \bigwedge_{j \in f(i)} \lnot \modobj_j o$ \\
    $\acc_i$ anti-preserves $\obfun_j$ for some $j \in f(i) \subseteq \ag$ & $w \acc_i u \imply \bigcap_{j \in f(i)} \obfun_j(u) \subseteq \obfun_i(w)$ & $\lnot \modobj_i o \to \square_i \bigvee_{j \in f(i)} \lnot \modobj_j o$  \\
    $\acc_i$ inverts $\obfun_j$ for all $j \in f(i) \subseteq \ag$         & $w \acc_i u \imply \obfun_i(w) \cap \bigcup_{j \in f(i)} \obfun_j(u) = \emptyset$ & $\modobj_i o \to \square_i \bigwedge_{j \in f(i)} \lnot \modobj_j o$ \\
    $\acc_i$ inverts $\obfun_j$ for some $j \in f(i) \subseteq \ag$        & $w \acc_i u \imply \obfun_i(w) \cap \bigcap_{j \in f(i)} \obfun_j(u) = \emptyset$ & $\modobj_i o \to \square_i \bigvee_{j \in f(i)} \lnot \modobj_j o$  \\
    $\acc_i$ anti-inverts $\obfun_j$ for all $j \in f(i) \subseteq \ag$    & $w \acc_i u \imply \obfun_i(w) \cup \bigcap_{j \in f(i)} \obfun_j(u) = \obj$ & $\lnot \modobj_i o \to \square_i \bigwedge_{j \in f(i)} \modobj_j o$ \\
    $\acc_i$ anti-inverts $\obfun_j$ for some $j \in f(i) \subseteq \ag$   & $w \acc_i u \imply \obfun_i(w) \cup \bigcup_{j \in f(i)} \obfun_j(u) = \obj$ &  $\lnot \modobj_i o \to \square_i \bigvee_{j \in f(i)} \modobj_j o$ \\
    \bottomrule
  \end{smallctabular}
  \caption{Some group properties.}
  \label{tbl:groproperties}
\end{table}

\medskip

The following proposition justifies the parametrisation of the group properties. In awareness epistemic terms, the first bullet says that, when combined with knowledge (or any other factive epistemic attitude), preservation and anti-preservation together imply that every agent is aware of the same things, and that this is common knowledge among all agents. This is clearly a trivialisation. The second bullet shows that knowledge cannot be combined with the universal group 
version of inversion or anti-inversion.

\begin{proposition}\label{prop:fsat}
  Let $f_{\mathit{gen}}=\{(i,\ag)\mid i \in \ag\}$, let $\nmod$ be a reflexive $\obj$-model.
  \begin{itemize}
    \item If $\nmod$ $f_{\mathit{gen}}$-satisfies universal preservation or anti-preservation, then all agents have available the same objects at each pair of worlds $w,v\in \wset$ connected by the transitive closure of $\bigcup_{i \in \ag} \acc_i$.

    \item $\nmod$ $f_{\mathit{gen}}$-satisfies neither universal inversion nor universal anti-inversion.
  \end{itemize}
\end{proposition}

\begin{rem}\label{rem:indgro}
  The individual version of (anti-)preservation and (anti-)inversion properties for $i \in \ag$ are the group versions (both universal and existential) for $f_{\mathit{indv}} = \{(i,\{i\})\mid i \in Dom(f)\}$.
\end{rem}

Finally, we can characterise the group properties using $\lanset$.

\begin{proposition}\label{pro:group-formulas}
  Let $f:\ag \to \wp(\ag)\setminus\{\emptyset\}$ be as described above; let $\nprop^{f}_{i}$ be any of the group properties of the left-hand column of \autoref{tbl:groproperties} (e.g., anti-inversion for $i$ and someone in $f(i)$) and let $\varphi(\nprop^{f}_{i})$ be its corresponding schema in the right-hand column. Let $\Gamma(\nprop^{f})$ the set of all instances of $\varphi(\nprop^{f}_{i})$ for all $i\in \ag$, and let $\nmod$ be an $\obj$-model. Then,
  \begin{center}
    $\nmod$ $f$-satisfies $\nprop$ \quad iff \quad  $\nmod \models \Gamma(\nprop^{f})$.
  \end{center}
\end{proposition}

Finally, here is the definition of the class of $\obj$-models satisfying a collection of properties.

\begin{definition}[Classes of models]\label{def:classesofmodels} 
  Let $(f_1,\ldots,f_n)$ be a sequence with $f_k: \ag \to \wp(\ag)\setminus\{\emptyset\}$ being a function as described above for every $1\leq k \leq n$, and let $(\nprop_1,\ldots,\nprop_n)$ be a sequence of group properties. We denote as $\mathfrak{M}({f_1\text{-}\nprop_1,\ldots,f_n\text{-}\nprop_n})$ the class of all $\obj$-models $\nmod$ s.t.\ for every $k$, $\nmod$ $f_k$-satisfies $\nprop_k$.
\end{definition}

\subsection{Axiom system}

Axiomatizing validities over $\classobmod$ (the class of all $\obj$-models) is straightforward, as formulas with the `owning' modality $\modobj_i$ can be seen as a particular atoms connected to a dedicated valuation function $\obfun_i$. Since the $\obfun_i$ sets have no particular requirements, the modal logic axiomatisation is enough.

\smallskip

When the focus is the class of models satisfying a certain collection of properties, additional work is needed; for this, \autoref{pro:group-formulas} will be useful. Define the notion of local semantic consequence w.r.t.\ a given class of models in the standard way \cite{blackburn2002}, denoting it by $\Phi \models_{\mathfrak{M}({f_1\text{-}\nprop_1,\ldots,f_n\text{-}\nprop_n})} \varphi$.

\begin{table}
  \begin{center}
    \begin{tabular}{r@{\;\;}l@{\qquad}r@{\;\;}l}
      \toprule
      TAUT: & All propositional tautologies                                            & MP: & From $\varphi$ and $\varphi \to \psi$, infer $\psi$\\
      K:    & $\square_i(\varphi \to \psi) \to (\square_i \varphi \to \square_i \psi)$ & NEC: & From $\varphi$ infer $\square_i \varphi$ \\
      \bottomrule
    \end{tabular}
  \end{center}
  \caption{The minimal modal logic $\mathsf{K}$.}
  \label{tab:staticaxioms}
\end{table}

\begin{definition}[Static logics]\label{def:static:logics}
  The logic $\mathsf{K}$ is the smallest set containing all instances of the axiom schemas of \autoref{tab:staticaxioms} that is moreover closed under both inference rules of the same table. The extension of $\mathsf{K}$ by $\Phi \subseteq \lanset$ is the smallest set of formulas containing all instances of schemas of \autoref{tab:staticaxioms}, all formulas in $\Phi$ and it is closed under both inference rules. Let $(f_1,\ldots,f_n)$ be a sequence of functions $\ag \to \wp(\ag)\setminus\{\emptyset\}$ as described above, and let $(\nprop_1,\ldots,\nprop_n)$ be a sequence of group properties. Then, we denote by $\nlog(f_1\text{-}\nprop_1,\ldots,f_n\text{-}\nprop_n)$ the extension of $\kripke$ with $\bigcup_{1 \leq k \leq n} \Gamma(\nprop^{f_k}_k)$.\footnote{See propositions \ref{pro:individual-formulas} and \ref{pro:group-formulas} for the meaning of $\Gamma(\nprop^{f})$.} Note that when $n=0$, $\nlog(f_1\text{-}\nprop_1,\ldots,f_n\text{-}\nprop_n)=\kripke$.
\end{definition}

The notions of $\nlog(f_1\text{-}\nprop_1,\ldots,f_n\text{-}\nprop_n)$-proof 
and $\nlog(f_1\text{-}\nprop_1,\ldots,f_n\text{-}\nprop_n)$-deduction from assumption (noted $\Phi \vdash_{\nlog(f_1\text{-}\nprop_1,\ldots,f_n\text{-}\nprop_n)}\varphi$),  are the standard ones in modal logic (see e.g., \cite{blackburn2002}).

\begin{theorem}[Static completeness]\label{thm:staticcompleteness} Let $(f_1,\ldots,f_n)$ be a sequence of functions $\ag \to \wp(\ag)\setminus\{\emptyset\}$ as described above, and let $(\nprop_1,\ldots,\nprop_n)$ be a sequence of group properties, we have that:

\begin{itemize}
\item[] $\nlog(f_1\text{-}\nprop_1,\ldots,f_n\text{-}\nprop_n)$ is sound and strongly complete with respect to $\mathfrak{M}({f_1\text{-}\nprop_1,\ldots,f_n\text{-}\nprop_n})$.
\end{itemize}
\end{theorem}

\section{Dynamics of \texorpdfstring{$\obj$}{O}-models, semantically}\label{sec:dynamics}

Changes in different modal attitudes (knowledge, beliefs, preferences and so on) have been the main topic of DEL. The main feature that distinguishes DEL from other approaches for modelling dynamics (e.g., propositional dynamic logic \cite{harel2001dynamic} or automata theory \cite{HopcroftMotwaniUllman2003}) is that changes are not represented as (binary) relations, but rather as operations that modify the underlying semantic structure. Indeed, DEL can be understood, more broadly, as the study of modal logics of model change \cite{sep-dynamic-epistemic}.
Here we focus on the \emph{event models} of \cite{baltag1998logic,baltag2004logics}: structures that, when `applied' to a relational model (by means of a \emph{product update}), produce another relational model. They were initially introduced as a way of modelling non-public acts of communication, and have since then been widely employed to model other forms of informational and factual changes \cite{baltag2008,van2008semantic, van2010dynamics,van2012action}. Besides their versatility, they have an important technical advantage: as proved in \cite{van2008semantic}, any pointed relational model can be turned into any other by means of the product update with some event model that allows factual change.\footnote{Slightly more precisely, given pointed models $(\nmod, w)$ and $(\nmod', w')$, there is `almost always' an event model such that, when applied to $(\nmod, w)$, produces a pointed model $(\nmod'', w'')$ that is, from the point of view of the language of propositional dynamic logic \cite{harel2001dynamic} (an extension of the basic modal language), indistinguishable from $(\nmod', w')$. See \cite{van2008semantic} for details.}
The rest of this section will discuss an extension of these event models that works for describing dynamics of $\obj$-models.

\begin{definition}[Event $\obj$-Model]
  An \emph{event $\obj$-model} is a tuple $\nemod=\emod$ where $\eveset\neq \emptyset$ is a finite set of events, $\T: \ag \to \wp(\eveset \times \eveset)$ assigns to each agent a binary relation, $\pre: \eveset \to \lanset$ assigns a precondition to each event, and $\eff: (\ag \times \{+,-\} \times \eveset) \to \wp(\obj)$ is a function indicating, for each event, its (positive and negative) effects on the set of objects available to each agent (write $\eff (i,\pm, s) $ as $\eff_i^{\pm}(s)$ for $\pm\in \{+,-\}$). We assume that, for every $s\in \eveset$ and every $i\in \ag$, the sets $\eff_i^{+}(s)$ and $\eff_i^{-}(s)$ are finite and disjoint. Note: $\T(i)$ abbreviates $\T_i$. The set of events of a given $\nemod$ is referred to as $\eveset[\nemod]$ (and the same convention applies for the other components of $\nemod$). We use infix notation for each $\T_i$. A \emph{pointed event $\obj$-model} is a tuple $(\nemod, s)$ with $\nemod=\emod$ an event $\obj$-model and $s \in \eveset[\nemod]$.
\end{definition}

The above definition does not include the \emph{post-condition function} (see, e.g., \cite{van2012action,tark2013eveawr}), as we want to focus on non-factual changes (i.e., changes on accessibility relations and $\obj$-sets, but not on atomic valuations). We think, however, that incorporating them does not pose any challenge, {since our framework can in fact be seen as a variation of event models for factual change, where one deals with agent-indexed predicates instead of purely atomic propositions.}

\begin{definition}[Product update]
  Let $\nmod=\objmod$ be an $\obj$-model and let $\nemod=\emod$ be an event $\obj$-model. The \emph{product update} of $\nmod$ and $\nemod$ produces the model $\nmod \otimes \nemod=(\wset',\acc',\obfun', \valfun')$, where:
  \vspace{-1ex}
  \begin{multicols}{2}
    \setlength{\leftmargini}{1.25em}
    \begin{itemize}
      \item $\wset':=\{ (w,s)\in \wset\times \eveset \mid \nmod,w\models \pre(s)\}$.
      \item $\acc_i':=\{\big ( (w,s),(u,t) \big )\in \wset'\times\wset' \mid w{\acc_i}u \;\&\; s{\T_i}t \}$.
      \item $\obfun'_i(w,s):=\big(\obfun_i(w)\cup \eff^{+}_i(s)\big)\setminus \eff^{-}_i(s)$.
      \item $\valfun'(p) :=\{(w,s)\in \wset' \mid w \in \valfun(p)\}$.
    \end{itemize}
  \end{multicols}
  \vspace{-1ex}
  \noindent Note: $\wset'$ is empty (and thus $\nmod \otimes \nemod$ is not defined) when no possible world satisfies any precondition. Thus, $\otimes$ is a partial function. When $\wset' \neq \emptyset$, we say that $\nmod \otimes \nemod$ is \emph{defined}.
\end{definition}

\msparagraph{The closure problem} Given a class of $\obj$-models $\mathfrak{M}$, the \emph{closure} problem \cite{aucher2008consistency,balbiani2012some} asks to find a class of event $\obj$-models $\mathfrak{E} \neq \emptyset$ s.t., $\nmod \in \mathfrak{M}$ and $\nemod \in \mathfrak{E}$ imply $\nmod\otimes \nemod \in \mathfrak{M}$. This is not trivial for the properties in Tables \ref{tbl:properties} and \ref{tbl:groproperties}: it is clear that executing certain event $\obj$-models in certain $\obj$-models leads to the loss of, e.g., individual preservation. 
This paper focusses on group properties \autorefp{rem:indgro}, using $\emp(\nprop)$ for referring to the event-model property in \autoref{tbl:indformproperties} that corresponds to the group property $\nprop$ in \autoref{tbl:groproperties}.\footnote{For instance, if $\nprop$ is anti-inversion for someone, then $\emp(\nprop)=\emp^{\mathsf{anti-inversion-}\exists}$.} 

\smallskip

\begin{table}[t]\label{tab:propertiesofevents}
\renewcommand{\arraystretch}{1.2}
  \begin{smallctabular}{@{}l@{\;\;}l@{\quad}l@{}}
    \toprule
    \textbf{$\nemod$ $f$-satisfies}
    &
    \textbf{iff for every $i\in Dom(f)$, $s,t \in \eveset[\nemod]$}
    &
    \textbf{is safe for}
    \\
    \midrule
    ${\emp^{\mathsf{pres-}\forall}}$           & $s\T_i t \imply \eff_i^{+}(s) \subseteq \bigcap_{j \in f(i)} \eff_j^{+}(t)$ and $ \bigcup_{j \in f(i)}\eff_j^{-}(t) \subseteq \eff^{-}_i(s)$ & preservation for everyone\\
    ${\emp^{\mathsf{pres-}\exists}}$           & $s\T_i t \imply \eff_i^{+}(s) \subseteq \bigcup_{j \in f(i)} \eff_j^{+}(t)$ and $ \bigcup_{j \in f(i)}\eff_j^{-}(t) \subseteq \eff^{-}_i(s)$ & preservation for someone\\
    ${\emp^{\mathsf{anti-pres-}\forall}}$      & $s\T_i t \imply \bigcup_{j \in f(i)} \eff_j^{+}(t) \subseteq \eff_i^{+}(s)$ and $\eff_i^{-}(s) \subseteq \bigcap_{j \in f(i)} \eff^{-}_j(t)$ & anti-preservation for everyone\\
    ${\emp^{\mathsf{anti-pres-}\exists}}$  & $s\T_i t \imply \bigcup_{j \in f(i)} \eff_j^{+}(t) \subseteq \eff_i^{+}(s)$ and $\eff_i^{-}(s) \subseteq \bigcup_{j \in f(i)} \eff^{-}_j(t)$ & anti-preservation for someone\\
   ${\emp^{\mathsf{inv-}\forall}}$             & $s\T_i t \imply \eff_i^{+}(s) \subseteq \bigcap_{j \in f(i)} \eff_j^{-}(t)$ and $\bigcup_{j \in f(i)} \eff_j^{+}(t) \subseteq \eff^{-}_i(s)$ & inversion for everyone\\
   ${\emp^{\mathsf{inv-}\exists}}$        &  $s\T_i t \imply \eff_i^{+}(s) \subseteq \bigcup_{j \in f(i)} \eff_j^{-}(t)$ and $\bigcup_{j \in f(i)} \eff_j^{+}(t) \subseteq \eff^{-}_i(s)$ & inversion for someone\\
   ${\emp^{\mathsf{anti-inv-}\forall}}$        & $s\T_i t \imply \eff_i^{-}(s) \subseteq \bigcap_{j \in f(i)} \eff_j^{+}(t)$ and $\bigcup_{j \in f(i)} \eff_j^{-}(t) \subseteq \eff^{+}_i(s)$ & anti-inversion for everyone\\
   ${\emp^{\mathsf{anti-inv-}\exists}}$        & $s\T_i t \imply \eff_i^{-}(s) \subseteq \bigcup_{j \in f(i)} \eff_j^{+}(t)$ and $\bigcup_{j \in f(i)} \eff_j^{-}(t) \subseteq \eff^{+}_i(s)$ & anti-inversion for someone\\
    \bottomrule
  \end{smallctabular}
  \caption{Properties of event $\obj$-models.}
  \label{tbl:indformproperties}
\end{table}

\begin{definition}[Classes of event models] Let $(f_1,\ldots,f_n)$ be a sequence of functions $\ag\to \wp(\ag)\setminus\{\emptyset\}$ as described above, and let $(\emp_1,\ldots,\emp_n)$ be a sequence of group properties for event models (\autoref{tbl:indformproperties}). We denote as $\mathfrak{E}({f_1\text{-}\emp_1,\ldots,f_n\text{-}\emp_n})$ the class of all event $\obj$-models $\nemod$ s.t.\ for every $1\leq k \leq n$, $\nemod$ $f_k$-satisfies $\emp_k$.
\end{definition}

With properties of event models defined, here is the main result.

\begin{theorem}[Closure for group properties]\label{thm:closure}
  Let $f: \ag \to \wp(\ag)\setminus \{\emptyset\}$ be as described above. Let $\nmod$ be an $\obj$-model and $\nemod$ an event $\obj$-model s.t. $\nmod \otimes \nemod$ is defined. For any property $\nprop$ in \autoref{tbl:groproperties}, if $\nmod$ $f$-satisfies $\nprop$ and $\nemod$ $f$-satisfies $\emp(\nprop)$, then $\nmod \otimes \nemod$ $f$-satisfies $\nprop$.
\end{theorem}

\begin{example}[Different forms of forgetting]
  \autoref{thm:closure} helps to test the compatibility between the model of a notion/concept and the model of its dynamics: a single action might be modelled by different event models, and the choice might depend on the specific model requirements. As an example, and in the awareness context, consider an action through which agent $i$ becomes unaware of the atom $p$ without anybody else noticing it. In \cite{van2010dynamics}, this action corresponds to the event model $\mathsf{Pri}_{i}^{p}=(\{\bullet,\circ\}, \T, \pre, \eff)$ with $\T_i=\{(\bullet,\bullet), (\circ,\circ)\}$ and $\T_j=\{(\bullet,\circ),(\circ,\circ)\}$ for $j\neq i$, and with $\eff^{-}_i(\bullet)=\{p\}$ and $\eff^{-}_j(\bullet)=\eff^{\pm}_j(\circ)=\eff^{\pm}_i(\circ)=\emptyset$. When $\ag=\{1,2\}$ and $i = 1$, the event model can be represented as 
  
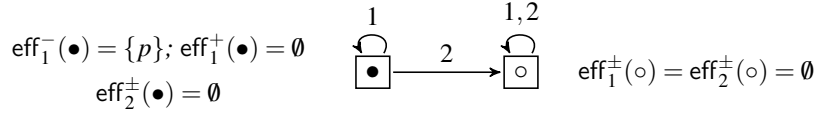
\begin{center}
    \begin{tikzpicture}[modal, world/.append style= {minimum size=1.5cm}]

    \node (w) [draw] {$\bullet$};
    \node (w1) [draw, right=1.5cm of w] {$\circ$};

    \draw[->] (w) edge[reflexive above] node[above]{\small $1$} (w);
    \draw[->] (w) edge node[above]{\small$2$} (w1);
    \draw[->] (w1) edge[reflexive above] node[above]{\small$1,2$} (w1);

    \node (aww) [left = 0em of w] {};
    \node (aww) [left = 0.05cm of aww, outer sep = 0] {\small \begin{tabular}{c}$\eff^{-}_1(\bullet)=\{p\}$; $\eff^{+}_1(\bullet)=\emptyset$\\[0.5em]$\eff^{\pm}_2(\bullet)=\emptyset$\end{tabular}};
    \node (12aww1) [right = 1em of w1] {\small $\eff^{\pm}_1(\circ)=\eff^{\pm}_2(\circ)=\emptyset$};
    \end{tikzpicture}
  \end{center}
 
  This event model does the job when awareness is not required to have special properties.\footnote{It even preserves the individual version of invariance (\autoref{tbl:properties}).} However, it is not appropriate, e.g., when $\acc$ is required to $f_{gen}$-anti-preserve $\obfun$ for everyone, for $f_{gen}=\{(i,\ag)\mid i \in \ag\}$ {(as in the case of awareness of arguments of \cite{schwarzentruber2012building,synthese})}.
  Fortunately, there is another event model that captures the central intuition of the action {(that is, that agent $1$ privately looses awareness of $p$ and she is the only one suffering this loss in the actual event $\bullet$)} while also preserving the property. Indeed, consider $\mathsf{AlPri}_{i}^{p}=(\{\bullet,\circ, \triangle\}, \T, \pre, \eff)$ with $\T_i=\{(\bullet,\triangle), (\triangle,\triangle),(\circ,\circ)\}$ and $\T_j=\{(\bullet,\circ), (\triangle,\triangle),(\circ,\circ)\}$ for $j\neq i$, and with $\eff^{-}_i(\bullet)=\eff^{-}_i(\triangle)=\eff^{-}_j(\triangle)=\{p\}$ and $\eff^{+}_i(\bullet)=\eff^{\pm}_j(\bullet)=\eff^{+}_i(\triangle)=\eff^{+}_j(\triangle)=\eff^{\pm}_i(\circ)=\eff^{\pm}_j(\circ)=\emptyset$. When $\ag=\{1,2\}$ and $i = 1$, the event model is
  
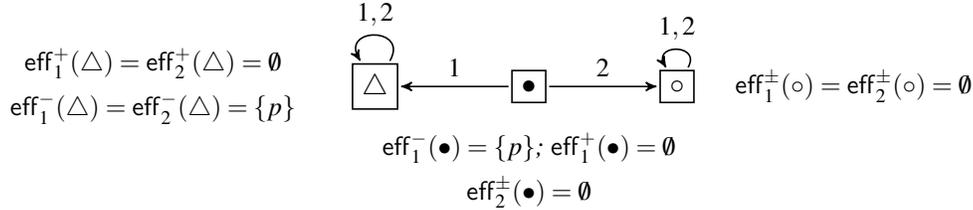
\begin{center}
    \begin{tikzpicture}[modal,world/.append style={minimum size=1.5cm}]
      \node (w) [draw] {$\bullet$};
      \node (w2) [draw, right=1.5cm of w]  {$\circ$};
      \node (w1) [draw, left=1.5cm of w]  {$\triangle$};
      
      \node (1aww) [below=0.25cm of w] {\small \begin{tabular}{c}$\eff^{-}_1(\bullet)=\{p\}$; $\eff^{+}_1(\bullet)=\emptyset$\\[0.5em]$\eff^{\pm}_2(\bullet)=\emptyset$\end{tabular}};
    
      \node (pos) [right=1.5cm of w] {};

      \draw[->] (w) edge node[above]{\small $1$} (w1);
      \draw[->] (w1) edge[reflexive above] node[above]{\small $1,2$} (w1);
      \draw[->] (w2) edge[reflexive above] node[above]{\small $1,2$} (w2);
      \draw[->] (w) edge node[above]{\small $2$} (w2);

      \node (12aw1) [left=0.4cm of w1] {\small \begin{tabular}{c}$\eff_1^{+}(\triangle)=\eff_2^{+}(\triangle)=\emptyset$\\[0.5em]$\eff_1^{-}(\triangle)=\eff_2^{-}(\triangle)=\{p\}$\end{tabular}};
      \node (12aw2) [right=0.4cm of w2] {\small $\eff_1^{\pm}(\circ)=\eff_2^{\pm}(\circ)=\emptyset$};
    \end{tikzpicture}
  \end{center}
  {Just as before, agent $1$ drops $p$ (the effect of $\bullet$), and this change is private, since $2$ believes that nothing happened ($\circ$). Additionally, and due to the nature of universal anti-preservation, $1$ thinks that everyone loses awareness of $p$ as well (the effects of $\bigtriangleup$). Note, moreover, that $\mathsf{AlPri}_{i}^{p}$ $f_{gen}$-satisfies $\emp^{\mathsf{anti-inv-}\forall}$ (our sufficient condition for the preservation of universal anti-preservation).}
\end{example}

\section{Dynamics of \texorpdfstring{$\obj$}{O}-models, syntactically}\label{sec:axioms}

Here is the language used to describe the effect of product updates.

\begin{definition}[Language $\lanset({\star})$]
  Let $\mathfrak{E}^{\obj}$ the class of all event $\obj$-models, and let $\star \subseteq \mathfrak{E}^{\obj}$ be a non-empty subclass. The dynamic language $\lanset({\star})$ is given by
  \begin{ctabular}{c}
    $\varphi ::= \top \mid p \mid \modobj_i o \mid \lnot \varphi \mid \varphi \land \varphi \mid \square_i \varphi \mid [\nemod,s] \varphi$ 
  \end{ctabular}
  with $p \in \at$, $i \in \ag$, $o \in \obj$, $\nemod \in \star$ and $s \in \eveset[\nemod] $. 
  The truth clause for the new kinds of formulas is:
  \begin{ctabular}{r@{\quad\text{iff}\quad}l}
    $\nmod,w \models [\nemod,s]\varphi$ & $\nmod,w\models \pre(s)$ implies $\nmod \otimes \nemod,(w,s)\models \varphi$.
  \end{ctabular}
\end{definition}  

\begin{definition}[Dynamic logics] Let $\nlog(f_1\text{-}\nprop_1,\ldots,f_n\text{-}\nprop_n)$ be a static logic of \autoref{def:static:logics}. The logic $\mathsf{L}^{!}(f_1\text{-}\nprop_1,\ldots,f_n\text{-}\nprop_n)$ extends $\nlog(f_1\text{-}\nprop_1,\ldots,f_n\text{-}\nprop_n)$ with all axioms and rules of \autoref{tab:redaxioms} that can be written in $\lanset(\mathfrak{E}(f_1\text{-}\emp(\nprop_1),\ldots,f_n\text{-}\emp(\nprop_n)))$.
\end{definition}

Then, the completeness result.

\begin{theorem}[Dynamic completeness]\label{thm:dycompleteness} Let $(f_1,\ldots,f_n)$ be a sequence of functions $\ag\to \wp(\ag)\setminus\{\emptyset\}$ as described above, and let $(\nprop_1,\ldots,\nprop_n)$ be a sequence of group properties. We have that:

\begin{itemize}
\item[] $\nlog^{!}(f_1\text{-}\nprop_1,\ldots,f_n\text{-}\nprop_n)$ is sound and strongly complete with respect to $\mathfrak{M}({f_1\text{-}\nprop_1,\ldots,f_n\text{-}\nprop_n})$.
\end{itemize}
\end{theorem}

\begin{table}
  \begin{center}
    \begin{tabular}{l@{\qquad}l}
      \toprule
      $[\nemod,s]\top\leftrightarrow \top$ & $[\nemod,s]\lnot \varphi \leftrightarrow (\pre(s)\to \lnot [\nemod,s] \varphi)$ \\
      $[\nemod,s] p \leftrightarrow (\pre(s)\to p)$ & $[\nemod,s] (\varphi \land \psi) \leftrightarrow ([\nemod,s] \varphi \land [\nemod,s] \psi)$ \\
      $[\nemod,s] \obj_i x \leftrightarrow (\pre(s)\to \obj_i x)$ \; for $x\notin \eff[\nemod]^{+}_i(s)\cup \eff[\nemod]^{-}_i(s)$ & $[\nemod,s] \square_i \varphi  \leftrightarrow (\pre(s)\to \bigwedge_{s\T_i t}\square_i[\nemod,t]\varphi)$ \\
      $[\nemod,s] \obj_i x \leftrightarrow \top$ \; for $x\in \eff[\nemod]^{+}_i(s)$ & \\
      $[\nemod,s] \obj_i x \leftrightarrow \lnot \pre(s)$ \qquad for $x\in \eff[\nemod]^{-}_i(s)$ & From $\varphi \leftrightarrow \psi$, infer $\delta \leftrightarrow \delta[\varphi/ \psi]$\\
      \bottomrule
    \end{tabular}
  \end{center}
  \caption{Reduction axioms for dynamic modalities. $\delta[\varphi/ \psi]$ is the result of substituting one or more occurrences of $\varphi$ in $\delta$ by $\psi$. Furthermore, it is assumed for simplicity that these substitutions do not affect the occurrences of formulas inside dynamic modalities, i.e. $([\nemod,s]\delta)[\varphi/\psi]=[\nemod,s](\delta [\varphi/\psi])$.}
  \label{tab:redaxioms}
\end{table}

\section{Conclusion and future work}\label{sec:conclusion}

{
The paper provides an abstract look at awareness epistemic models, understanding them not as a representation of the formulas the agents are aware of, but rather as a more general setting for dealing with a notion of `owning abstract objects'. As discussed in \autoref{sec:basic}, several well-know proposals from different areas can be seen as particular instances of these general type of structures.

When modelling specific phenomena, a general $\obj$-structure may be asked to satisfy specific requirements. Of particular interest are those that relate $\obfun$-sets with accessibility relations, and \autoref{sbs:properties} listed some possibilities, together with their characterising formula. Maybe more importantly, these requirements should be preserved by model operations representing dynamics of the modelled phenomena. \autoref{sec:dynamics} focussed on model operations defined in terms of event models, introducing classes of the latter that, under the product update operation, guarantee the preservation of the specified requirements. This establishes a form of `compatibility' between the represented phenomena and the chosen event models. \autoref{sec:axioms} closed the discussion, obtaining complete axiomatisations via reduction axioms.

There are branches open for further exploration; here are two of them. The first one is to work out the details of the instantiations of $\obj$-models that were sketched in \autoref{sec:basic}. The second one is to study more systematically the trivialisation of awareness ($\obfun$-sets) for extreme cases of $f$ (e.g., for $f_{gen}$) so as to underpin our notion of $f$-satisfiability.
}

\bibliographystyle{eptcs}
\bibliography{tesisC}

\section*{Appendix}\label{sec:appendix}

\msparagraph{\autoref{thm:staticcompleteness}} {Let $(f_1,\ldots,f_n)$ be a sequence of functions $\ag \to \wp(\ag)\setminus\{\emptyset\}$ as described above, and let $(\nprop_1,\ldots,\nprop_n)$ be a sequence of group properties, we have that:

\begin{itemize}
\item[] $\nlog(f_1\text{-}\nprop_1,\ldots,f_n\text{-}\nprop_n)$ is sound and strongly complete with respect to $\mathfrak{M}({f_1\text{-}\nprop_1,\ldots,f_n\text{-}\nprop_n})$.
\end{itemize}}
\begin{proof}
Let $\nlog(f_1\text{-}\nprop_1,\ldots,f_n\text{-}\nprop_n)$ and $\mathfrak{M}(f_1\text{-}\nprop_1,\ldots,f_n\text{-}\nprop_n)$ be arbitrarily fixed from now on, we drop the parameters $(f_1\text{-}\nprop_1,\ldots,f_n\text{-}\nprop_n)$ for readability. \par 
Soundness follows by induction for the length of $\nlog$-proofs. For the basic step, one needs to show that every instance of an $\nlog$-axiom schema is valid in the corresponding class of models. For the inductive step, it is enough to show that both inference rules preserve $\mathfrak{M}$-validity. \par 
As for completeness, the proof follows a canonical model argument.  
We denote by $\mathsf{MC}^{\nlog}$ the class of all maximally $\nlog$-consistent sets of formulas. The proofs of the Lindenbaum lemma, as well as the closure properties of maximally $\nlog$-consistent sets, are as usual. The $\nlog$-canonical model is the $\obj$-model $\nmod^{\nlog}=(\wset^{\nlog},\acc^{\nlog},\obfun^{\nlog},\valfun^{\nlog})$ where each component is defined as follows:

\begin{center}
\begin{tabular}{r l}

$\wset^{\nlog}=$ & $ \mathsf{MC}^{\nlog}\text{,}$\\
$\Phi \acc^{\nlog}_{i}\Delta$ \quad iff & $\{\varphi \in \lanset \mid \square_i \varphi \in \Phi\}\subseteq \Delta$, \\
$\obfun_i^{\nlog}(\Phi)=$& $\{x \in \obj \mid \obj_i x \in \Phi\}$, and \\
$\valfun^{\nlog}(p)=$ &$\{\Phi\in \wset^{\nlog} \mid p \in\Phi \}$.
\end{tabular}
\end{center}

The proof of the Truth Lemma ($\forall \varphi \in \lanset$, $\varphi \in \Phi$ iff $\nmod^{\nlog},\Phi \models \varphi$) is by induction on the structure of $\varphi$. The only difference w.r.t.\ the proof of the lemma for basic modal logic is the step where $\varphi=\obj_i x$, and this is straightforward. For the right-to-left direction of the case $\varphi=\square_i \psi$, one needs to show that the so-called Existence Lemma holds, namely, that if $\lnot \square_i \delta \in \Phi (\in \wset^{\nlog}) $, then there is a $\Delta \in \wset^{\nlog}$ with $\Phi \acc^{\nlog}_i \Delta$ and $\lnot \delta \in \Delta$.
The final, crucial part is to show that all group properties are canonical, that is to say, if $\varphi^{f}$ is an $\nlog$-axiom schema that defines a group property $\nprop$, then $\nmod^{\nlog}$ $f$-satisfies $\nprop$. We leave details for the reader.

As a curiosity, note that if $\obj$-formulas are considered as special types of atoms (as done, e.g., in \cite{synthese}), our logic is not normal in the sense of \cite{blackburn2002}, because the rule of uniform substitution does not preserve validity in all classes of models. However, this does not affect the completeness argument.
\end{proof}

\nsparagraph{\autoref{thm:closure}}
{Let $f: \ag \to \wp(\ag)\setminus \{\emptyset\}$ be as described above. Let $\nmod$ be an $\obj$-model and $\nemod$ an event $\obj$-model s.t. $\nmod \otimes \nemod$ is defined. For any property $\nprop$ in \autoref{tbl:groproperties}, if $\nmod$ $f$-satisfies $\nprop$ and $\nemod$ $f$-satisfies $\emp(\nprop)$, then $\nmod \otimes \nemod$ $f$-satisfies $\nprop$.}
\begin{proof}  
{For space reasons, we just show that the theorem holds for the first and the last property. The rest of the cases are left for the reader:}

\noindent \textbf{[$\nprop=$ preservation for everyone]} Take $\nmod$ and $\nemod$ s.t.\ 
\begin{ctabular}{r l}
$\nmod$ $f$-satisfies preservation for everyone & (1) \\
$\nemod$ f-satisfies ${\emp^{\mathsf{pres-}\forall}}$ & (2)
\end{ctabular}
We want to show that $\nmod \otimes \nemod=(\wset',\acc',\valfun',\obfun')$ $f$-satisfies preservation for everyone. Let $i\in Dom(f)$ and $(w,s)\in \wset'$ and suppose that $(w,s)\acc'_i (u,t)$. This is equivalent by the definition of product update to 
\begin{ctabular}{r l}
$w\acc_i u$ \quad and \quad  $s\T_i t$ & (3) 
\end{ctabular}
Further, suppose that $x \in \obfun'_i(w,s)$, which is equivalent, by the definition of product update, to $x \in \big(\obfun_i(w)\cup \eff^{+}_i(s)\big)\setminus \eff^{-}_i(s)$. We continue by cases on the membership of $x$, showing that $x \in \bigcap_{j \in f(i)}\obfun'_j(u,t)$ always obtains.
\par 
\textbf{Case: $x \in \obfun_i(w)$ and $x\notin \eff^{-}_i(s)$.} On the one hand, $x\in \obfun_i(w)$ implies together with (1) and (3) that $x \in \bigcap_{j \in f(i)} \obfun_j(u)$. On the other hand, $x\notin \eff^{-}_i(s)$ implies together with (2) and (3) that $x \notin \bigcup_{j\in f(i)}\eff^{-}_j(t)$. Both facts imply by set-theoretic reasoning that $x \in \bigcap_{j \in f(i)}\big((\obfun_j(u)\cup \eff^{+}_j(t))\setminus\eff^{-}_j(t)\big)$, which is equivalent to what we wanted to show (by definition of product update).
\par 
\textbf{Case: $x \in \eff^{+}_i(s)$ and $x\notin \eff^{-}_i(s)$.} The latter implies, together with (2) and (3), that $x \in \bigcap_{j \in f(i)}\eff^{+}_j(t)$ and $x\notin \bigcup_{j \in f(i)}\eff^{-}_j(t)$, which implies by set-theoretic reasoning that $x \in \bigcap_{j \in f(i)}\big((\obfun_j(u)\cup \eff^{+}_j(t))\setminus\eff^{-}_j(t)\big)$, which is equivalent to what we wanted to show (by definition of product update).
 \medskip

\noindent \textbf{[$\nprop=$ anti-inversion for someone]} Take $\nmod$ and $\nemod$ s.t.\ 
\begin{ctabular}{r l}
$\nmod$ $f$-satisfies anti-inversion for someone & (1) \\
$\nemod$ f-satisfies ${\emp^{\mathsf{anti-inv-}\exists}}$ & (2)
\end{ctabular}
We want to show that $\nmod \otimes \nemod=(\wset',\acc',\valfun',\obfun')$ $f$-satisfies anti-inversion for someone. Let $i\in Dom(f)$ and $(w,s)\in \wset'$ and suppose that $(w,s)\acc'_i (u,t)$. This is equivalent by the definition of product update to 
\begin{ctabular}{r l}
$w\acc_i u$ \quad and \quad  $s\T_i t$ & (3). 
\end{ctabular}
Further, suppose that $x \notin \obfun'_i(w,s)$, which is equivalent, by the definition of product update, to $x \notin \obfun_i(w)\cup \eff^{+}_i(s)\setminus \eff^{-}_i(s)$. We want to show $x \in \bigcup_{j \in f(i)}\obfun'_j(u,t)$, which is equivalent to $x \in \bigcup_{j \in f(i)}(\obfun_j(u)\cup \eff^{+}_j(t))\setminus \eff^{-}_j(t)$. We continue by cases on $x \notin \obfun_i(w)\cup \eff^{+}_i(s)\setminus \eff^{-}_i(s)$, showing that the latter claim always obtains.

\textbf{Case: $x \notin \obfun_i(w)$ and $x\notin \eff^{+}_i(s)$.} On the one hand, $x\notin \obfun_i(w)$ implies together with (1) and (3) that $x \in \bigcup_{j \in f(i)}\obfun_j(u)$. On the other hand, $x\notin \eff^{+}_i(s)$ implies together with (2) and (3) that $x \notin \bigcup_{j \in f(i)}\eff^{-}_j(t)$. Both facts imply by set-theoretic reasoning that $x \in \bigcup_{j \in f(i)}(\obfun_j(u)\cup \eff^{+}_j(t))\setminus \eff^{-}_j(t)$.
 
\textbf{Case: $x \in \eff^{-}_i(s)$.} The latter implies, together with (2) and (3), that $x \in \bigcup_{j \in f(i)} \eff^{+}_j(t)$ which implies that $x \notin \bigcap_{j \in f(i)}\eff^{-}_i(t)$ (by definition of event $\obj$-model, because $\eff_k^{+}(t)\cap \eff^{-}_k(t)=\emptyset$ for every $k\in \ag$). The latter two claims implies by set-theoretical reasoning that $x \in \bigcap_{j \in f(i)}(\obfun_j(u)\cup \eff^{+}_j(t))\setminus \eff^{-}_j(t)$.
\end{proof}

\msparagraph{\autoref{thm:dycompleteness}}
{Let $(f_1,\ldots,f_n)$ be a sequence of functions $\ag\to \wp(\ag)\setminus\{\emptyset\}$ as described above, and let $(\nprop_1,\ldots,\nprop_n)$ be a sequence of group properties. We have that:

\begin{itemize}
\item[] $\nlog^{!}(f_1\text{-}\nprop_1,\ldots,f_n\text{-}\nprop_n)$ is sound and strongly complete with respect to $\mathfrak{M}({f_1\text{-}\nprop_1,\ldots,f_n\text{-}\nprop_n})$.
\end{itemize}}
\begin{proof}[Sketched]
Let $\mathfrak{M}(f_1\text{-}\nprop_1,\ldots,f_n\text{-}\nprop_n)$, $\lanset(\mathfrak{E}(f_1\text{-}\emp(\nprop_1),\ldots,f_n\text{-}\emp(\nprop_n)))$, $\nlog^{!}(f_1\text{-}\nprop_1,\ldots,f_n\text{-}\nprop_n)$, and $\nlog(f_1\text{-}\nprop_1,\ldots,f_n\text{-}\nprop_n)$ be arbitrarily fixed from now on. We drop the parameters and denote them by $\mathfrak{M}$, $\lanset(\mathfrak{E})$, $\nlog^{!}$, and $\nlog$ for the sake of readability, {but note that the parametrisation of each of the components is crucial for our argument.}

The soundness of $\nlog^{!}$ follows from soundness of its static base $\nlog$ (\autoref{thm:staticcompleteness}), the validity of axioms of \autoref{tab:redaxioms}, and the validity-preserving character of the only rule present in the same table. For proving the latter, i.e., that the application of the rule preserves validity within $\mathfrak{M}$, \autoref{thm:closure} is necessary. In more detail, the validity-preservation of the rule is proven by induction on $\delta$, and \autoref{thm:closure} is crucial when we arrive at the step where $\delta$ has the shape $[\nemod,s]\alpha$. {Moreover, and in the same inductive step, the simplification shown in the caption of Table \ref{tab:redaxioms} is needed.}

\par We can then prove strong completeness via a reduction argument (see \cite{kooi2007expressivity,hans2007,van2006logics,wang2013axiomatizations}). 
For doing so, we use two numeric measures for formulas, the \emph{depth} of $\varphi$, noted $d(\varphi)$, and the number of \emph{nested dynamic modalities} in $\varphi$, noted $Od(\varphi)$. More formally:

\begin{itemize}

\item Define $d:\lanset(\mathfrak{E}) \to \mathbb{N}$ as $d(p)=0$ for every $p \in \at$, $d(\obj_i x)=0$ for every $x \in \obj$, $i \in \ag$, $d(\circledast \varphi)=1+d(\varphi)$ where $\circledast \in \{\lnot, \square_i, [\nemod,s]\}$ and $d(\varphi\land \psi)=1+max(d(\varphi),d(\psi))$.

\item Define $Od:\lanset(\mathfrak{E})  \to \mathbb{N}$ as $Od(p)=0$, $d(\obj_i x)=0$ for every $x \in \obj$, $i \in \ag$, \quad $Od(\lnot \varphi)=Od(\square_i \varphi)= Od(\varphi)$,  $Od(\varphi\land \psi)=max(Od(\varphi),Od(\psi))$, and  $Od([\nemod,s]\varphi)=1+Od(\varphi)$.

 \end{itemize}

Now, we define the following function, translating formulas from each dynamic language $\lanset(\mathfrak{E})$ to the its static fragment $\lanset$:

\newcommand{\pam}{[\nemod,s]}
\newcommand{\paam}{[\mathcal{F},s]}
\begin{tabular}{l l}
 
$\tau (p)=p$ & $\tau([\nemod,s]p)=\pre(s)\to p $ \\
$\tau (\obj_i x) =\obj_i x$ & $\tau([\nemod,s]\obj_i x)=\pre(s) \to \obj_i x$ if $x\notin \eff[\nemod]^{+}_i(s)\cup \eff[\nemod]^{-}_i(s)$ \\
 & $\tau([\nemod,s]\obj_i x)=\top$ if $x\in \eff[\nemod]^{+}_i(s)$ \\
 & $\tau([\nemod,s]\obj_i x)=\lnot \pre(s)$ if $x\in \eff[\nemod]^{-}_i(s)$ \\
$\tau( \lnot \varphi)=\lnot \tau (\varphi)$ & $\tau([\nemod,s] \lnot \varphi)= \pre(s)\to \lnot \tau([\nmod,s]\varphi)$ \\
$\tau(\varphi \land \psi)=\tau(\varphi)\land \tau(\psi)$ & $ \tau([\nemod,s](\varphi \land \psi))=\tau(\pam \varphi) \land \tau(\pam\psi)$ \\
$\tau(\square_i \varphi)=\square_i\tau (\varphi)$ & $\tau(\pam\square_i \varphi)=\pre(s)\to \bigwedge_{s \T_i t} \square_i \tau([E,t]\varphi)$ \\

 & $\tau (\pam \paam \varphi)= \tau(\pam \tau(\paam \varphi))$ \\

\end{tabular}
\par \medskip
The next step is showing that the co-domain of $\tau$ is in fact $\lanset$. This is proven in two phases. First, one can show that it holds for the special case where $O(\varphi)=1$, and this is done by induction on $d(\varphi)$. Then it can be proven for the general case (and the previous claim is needed). Note that this translation amounts to what \cite{wang2013axiomatizations} coined as an \emph{inside-out} reduction because, when dealing with a formula $\delta$ with more that one nested dynamic operator (i.e., with $Od(\delta)\geq 2$), we first take care of the innermost occurrence due to the definition of $\tau$. \par 

Finally, we can establish and prove the key Reduction Lemma, namely, that for every $\varphi \in \lanset(\mathfrak{E})$: 
\begin{center}
$\vdash_{\nlog^{!}} \varphi \leftrightarrow \tau(\varphi)$.
\end{center}
This is done through a complex inductive argument. Again, one first needs to prove the claim for the special case $Od(\varphi)=1$ by induction on $\varphi$. Then, the claim can be proven for the general case. This second proof requires a double induction, first on $d(\varphi)$ and, we arrive at the step $\varphi=\pam\psi$, we continue by induction on $d(\psi)$. Note that the validity-preservation character of the rule of \autoref{tab:redaxioms} is strongly needed for all cases (which in turn requires \autoref{thm:closure}, as we mentioned).
\end{proof}
\end{document}